# Addressing the Challenge of Distributed Interactive Simulation With Data Distribution Service

*Akram HAKIRI*[1, 2],
*Pascal BERTHOU*[1, 2],
*Thierry GAYRAUD*[1,2]

[1]CNRS ; LAAS, 7, avenue du Colonel Roche, 31077 Toulouse, France
[2] Université Toulouse; UPS, INSA, INP, ISAE; LAAS; F-31077 Toulouse, France
Email: {Hakiri, Berthou, Gayraud}@laas.fr

Keywords:
HLA, RTI, DDS, QoS policy, Performance evaluation.

**ABSTRACT**: *Real-Time availability of information is of most importance in large scale distributed interactive simulation in network-centric communication. Information generated from multiple federates must be distributed and made available to interested parties and providing the required QoS for consistent communication. The remainder of this paper discuss design alternative for realizing high performance distributed interactive simulation (DIS) application using the OMG Data Distribution Service (DDS), which is a QoS enabled publish/subscribe platform standard for time-critical, data-centric and large scale distributed networks. The considered application, in the civil domain, is used for remote education in driving schools. An experimental design evaluates the bandwidth and the latency performance of DDS and a comparison with the High Level Architecture performance is given.*

## 1. Introduction

The High Level Architecture (HLA) [1] is the glue that allows the combination of computer simulations into large scale real-time live simulation that combines the air traffic control, logistic control and helps the reuse and the interoperability of distributed applications.

A promising approach to building and evolving large scale distributed simulation are standards-based QoS enabled publish/subscribe (pub/sub) platform that enable applications to communicate by publishing information they have and subscribing to information they need in timely manner. The recently adopted Data Distribution Service (DDS) [2] specification defines an application level interface that supports the Data-Centric Publish-Subscribe (DCPS) in real-time systems, mission, and safety critical application domain like defense, large scale networks and data conferencing applications.

This paper is twofold: 1) it describes the architecture of the OMG-DDS, which is a QoS enabled pub/sub platform standard, and (2) it evaluates the implementation of this architecture to investigate its design tradeoff and its performance and comparing it to HLA.

The remainder of this paper is organized as follow: after a brief introduction, Section 2 introduces the background and related work on Distributed Interactive Simulation (DIS) applications and drawn the limits of these solutions. Section 3 summarizes the DDS specification and its architectural overview. Section 4 describes the hardware configuration of our testbed and introduces the simulation design. Section 5 analyzes the results of experimentation. Conclusion and perspectives are given in last section

## 2. Overview of DIS standards

Distributed simulation aims at proposing a common architecture for communication allowing the integration and the interconnection of large scale simulators. Modeling and simulation (M&S) consists of techniques and tools for testing, analyzing and training in which real world and conceptual systems are reproduced by model. It allows the reducing of the time and the cost of the design of prototypes, their developments, their tests and the refinement of their life cycle. Moreover, it offers also practical means to evaluate the performances of the models.

After the success of SIMNET [3], DIS [4] was developed to address the interoperability of heterogeneous simulators. The essence of DIS is the creation of synthetic environment within which humans and simulations interact at multiple networked sites.

DIS was not fully distributed; each message must be received and treated by each node, which clutter the bandwidth even though not a lot of data is transmitted. DIS does not manage latency and causality that made the reusability of simulations impossible. Latencies were not controlled and no time management service was incorporated which caused data losses due to the rejection of too old packets.

Since many years, the standard of the DIS protocol has provided strong foundations for distributed real-time simulation. DIS was largely accepted by the industrials and the governments. ALSP Protocol was conceived to support simulation with discrete events and was implemented successfully in the sets of on line combat games [5]. To unify these fields and to extend their success towards the existing applications, the American Department of Defense (DOD) has proposed in 1995 the development of a new standard for modeling and simulation called HLA. HLA is an initiative to capture the best sides of DIS and ALSP and to provide at the same time a standard architecture for software simulation.

HLA is foremost a general purpose, reusable software architecture for the development and execution of very large distributed simulation application. The HLA has a wide applicability, across a full range of simulations areas, including education and training, analysis, engineering; web based distributed applications, real-time critical applications and variety of level resolution. Thus, the HLA supports interfaces to live participants, such as instruments platforms and live systems. These widely different applications areas indicate the variety of requirements that have been considered in development and evolution of the HLA.

In HLA terminology, a set of simulations that is capable of interoperating is a federation, and the individual simulations are federates (see figure 2). The HLA standard has three documented parts:

 - **Rules:** HLA-Rules are principals and conventions that must be followed to achieve proper interaction of federation during the federation execution. Five rules are related to the Federation Execution, where the others five rules are specific to the federate. HLA Rules ensure proper interactions of simulations in federation and describes the simulation and federate responsibilities.

 - **Object Model Template:** a formal model for specifying simulation data in term of hierarchy of object class, attributes, interactions and interaction parameters. HLA-OMT provides a common method for recording information and establishes the format of three key models: Federation Object Model (FOM), Simulation Object Model (SOM) and Management Object Model (MOM). Figure 1 shows a basic example of HLA-OMT.

```xml
<?xml version="1.0" encoding="UTF-8"?>
<objectModel
DTDversion="1516.2" name="Platsim.xml"
type="FOM" version="1.0" date="11-11-2009"
Auhter="Hakiri Akram" sponsor="LAAS-CNRS">
<objects>
    <objectClass name="Vehicule">
        <attribute name="VehiculeATT"
        transportation="HLAreliable"/>
    </objectClass>
</objects>
  <interactions>
    <interactionClass name="Global_Interaction">
      </interactionClass>
    </interactions>
</objectModel>
```

Figure 1: Example of the HLA-OMT

- **Interface Specification:** The interface specification (HLA-IS) is abstract; it aims to standardize an approach to persistent problem in distributed application. Services in both directions are defined as procedure call that take and return a parameter. The HLA-IS identifies how federates will interact with the federation, and ultimately with one another. It provides services and communication mechanism, forming a piece of software to ensure the information exchange usually implemented within Run-Time Infrastructure (RTI). There are six classes of services:

- *Federation Management services* offer basic functions required to create and operate a federation.
- *Declaration Management Services* include publication, subscription and supporting control functions. Federates which produce Object Class Attributes or Interaction must declare exactly what they are able to publish.
- *Object Management Services* involves registration, updates and dynamic transfer of the object and attributes.
- *Ownership Management Services* allow federates to distribute the responsibility for updating and deleting object instance and transfer the ownership of object/attributes.
- *Time Management Services* focus on the mechanics required to establish synchronization between distributed entities at runtime.

- *Data Distributed Management Services* provide a flexible and an efficient routing of data among federates for isolating publishers and subscribers.

RTI is a software which implements the interface specification of the HLA. It provides services in producer/consumer paradigm. RTI provides a C++ library (other languages like Java, C#, Ada exist), libRTI, through which federates can exchange data.

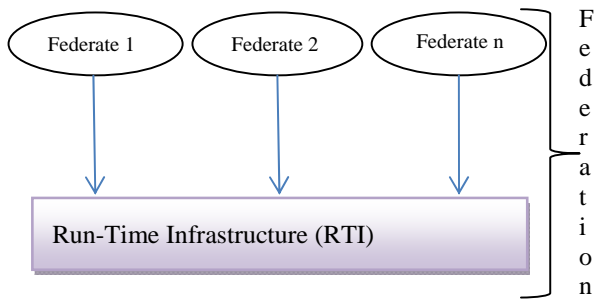

Figure 2: HLA Federation conceptual view, with federates exchanging data through the RTI.

Within *libRTI*, the class *RTIAmbassador* bundles the services provided by the RTI. All requests made by a federate on the RTI take the form of *RTIAmbassador* method call. The abstract *FederateAmbassador* identifies the callback functions each federate is obliged to provide.

## 3. Overview of Data Distributed Service

### 3.1. Core Features

Data Distribution Service (DDS) is a network middleware for distributed real-time application which simplifies application development, deployment and maintenance and provides fast, predictable distribution of real-time critical data over heterogeneous networks.

DDS Specification offers two levels of interface: one is a low level layer, the Data-Centric Publisher-Subscriber (DCPS), highly configurable, closely related to data and rich of QoS policies to determine the application required behavior. The Data Local Reconstruction Layer (DLRL) is the higher layer of the specification which is conceived to provide easy to use DCPS elements for developers. It summarizes the way to which an application can connect to DCPS through its proper classes using oriented Programming Object.

DLRL is an optional layer according to the OMG-DDS specifications.

The OMG DDS specifies a coherent set of profiles that target real time information-availability for domains ranging from small-scale embedded control systems up to large scale enterprise management systems. Each DDS-profile adds distinct capabilities that define the service level offered by DDS in order to realize this "right data at the right time at the right place paradigm":

- The *Minimum Profile* uses the publish/subscribe model to provide a high efficient information exchange between multiples publishers and subscribers in small area to large scale communication environment. This profile also involves the QoS policies that allow the middleware to match requested and offered QoS parameters.
- The *Ownership Profile* offers for replicated publishers the ability of the expression of fine grained specific information to interested parties.
- The *Content Subscription Profile* provides features to improve content filter information like those used in SQL language.
- The *Persistence Profile* offers transparent and tolerant durability of exchanged information.

Furthermore, DDS involves features which are designed to meet the needs of distributed real time applications: efficient data transfer with minimal latency, managing multiple source/sink of the same data; multiple independent communications networks (Domains) each using DDS can be used to over the same network transport protocol. Applications are only able to participate in the domains to which they belong, or it can be configured individually to participate in multiple domains.

DDS presents a virtual data space for sharing information (Global Data Space) between participants. Applications can read and/or write data objects addressed by the identifier field (Domain ID), the name of the Topic and a key.

The organization of the information exchange between distributed application is based on the Publish/subscribe (PS) System with the aid of the following constructs: Publisher and DataWriter on the Sending side, Subscriber and DataReader at the receiving side. Figure 3 illustrates the relationship between these objects.

- *Publisher*: is the object responsible for the actual sending of data. It owns and manages the DataWriter. An application uses DataWriters to send data. A DataReader can be only owned by a single Publisher while a Publisher can own many DataWriters.

- *Subscriber*: is the actual object responsible for the actual receipt of published data. The subscriber own and manages DataReaders. A DataReader can be only owned by a single subscriber while a subscriber can own multiple DataReaders.
- *Topic*: the association of DataWriter and DataReader is made by Topic. Topic associates a single name in the system (ID or Key), a type of data and the parameters of QoS specific to each data.
- *Domain*: It provides a Global Virtual Data Space where participants (Publisher/Subscriber) having the same Domain ID can exchange information. The Domain consists of several DomainParticipant which isolate participants into several sub domains.
- *DomainParticipant*: it is an entity which represents DDS application participation associated with the Domain. It serves as a container and manager to DDS entities.

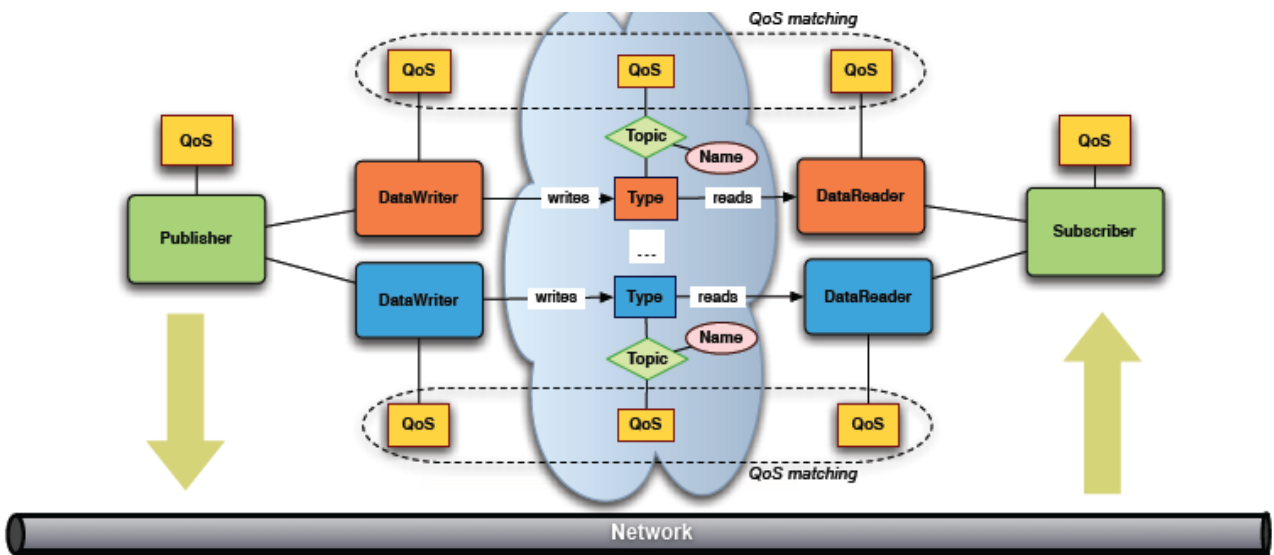

Figure 3 Architectural Overview of the DDS Architecture [8]

### 3.2. Benefits of DDS

DDS specifies transferred data as signals, streams and states. Signals characterize data that are continuously changing, so they are selected also as periodic traffic. Streams are snapshots of context previous sent data; they are selected as sporadic data. States describes the most current state of distributed entities, they do not change at fixed period and they can be elected as aperiodic data.

DDS addresses some new aspects not yet addressed by HLA, such as a rich set of QoS policies based on 'Request/Offered" contract including among others durability, liveliness, deadline, transport priority and more, while leaving out some other aspects addressed by HLA, such as time management and federation management. QoS policies provide a generic mechanism for the distributed applications to control the behavior of an entity [6, 7].

Figure 4 shows the QoS policies addressed by DDS: the first column specifies the QoS name, there are twenty QoS policies. Since QoS is comprised of individual QoS policies, it may be associated with a corresponding entity in the system, such as Topic (T), DataWriter (DW), DataReader (DR), DomainParticipant (DP), Publisher (P) or subscriber (S) (see column 2).

In several cases, for communication to occur efficiently, a QoS Policy on the publisher side must be compatible with a corresponding policy on the subscriber side. If the subscriber requests to receive data reliably while publisher defines a best-effort policy, communication will not happen as requested. To overcome this shortcoming, the subscriber and the publisher negotiate their QoS through Requested-Offered contract. In the pattern, the subscriber can specify a requested value for particular QoSPolicy (see column 3 in figure 4) to be set in compatible manner between the corresponding participants. An RxO setting of Yes (Y) indicates that policy can be set both at the publishing and subscribing side. Whereas if RxO is set to No (N) it indicates that the policy can be set in the two sides but the end settings are independents. Finally, if RxO is set to N/A (-) then compatibility does not apply.

| QoS Policy | Applicability | RxO | Modifiable | |
|---|---|---|---|---|
| DURABILITY | T, DR, DW | Y | N | Data Availability |
| DURABILITY SERVICE | T, DW | N | N | |
| LIFESPAN | T, DW | - | Y | |
| HISTORY | T, DR, DW | N | N | |
| PRESENTATION | P, S | Y | N | Data Delivery |
| RELIABILITY | T, DR, DW | Y | N | |
| PARTITION | P, S | N | Y | |
| DESTINATION ORDER | T, DR, DW | Y | N | |
| OWNERSHIP | T, DR, DW | Y | N | |
| OWNERSHIP STRENGTH | DW | - | Y | |
| DEADLINE | T, DR, DW | Y | Y | Data Timeliness |
| LATENCY BUDGET | T, DR, DW | Y | Y | |
| TRANSPORT PRIORITY | T, DW | - | Y | |
| TIME BASED FILTER | DR | - | Y | Resources |
| RESOURCE LIMITS | T, DR, DW | N | N | |
| USER_DATA | DP, DR, DW | N | Y | Configuration |
| TOPIC_DATA | T | N | Y | |
| GROUP_DATA | P, S | N | Y | |

Figure 4: QoS policies addressed by DDS [8]

The changeable property determines whether the QoSPolicy can be modifiable (see column 4 in figure 4) after the entity is enabled.

DDS-DCPS groups the several QoS Policies into concerning groups (see column 5 from figure 4). Users will employ the desired QoS policy to address the specific need of is application. I should be noted that Resources QoS Policy group can be mapped into the underlying network, for instance the QoS TRANSPORT_PRIORITY may be applied to the DiffServ Infrastructure in order to enumerate the CodePoint field.

Among the HLA services enumerated in Section 2, the Time Management Service is not supported within DDS. It was primarily specified for Parallel And Distributed Systems (PADS). The HLA standardized APIs specifies a save/restore services which ensure the creation of synchronization point between distributed systems to offer more consistence and reliability to applications.

In another hand, the HLA-RTI allows to applications to choose the degree to which it participate in time management. The Time Management has to do with ensuring that events are delivered to applications in correct order, but the order in which events arrive at the remote application cannot be guaranteed. Events do not arrive in the order of cause and effect.

Thus, the close difference between the HLA and DDS middleware may appears when evaluating their performance. But this does not prevent getting very well performance in several distributed applications.

DDS is key enabling technology and Next-Generation based Warfare Systems which deliver extremely high performance, high availability & reliability, along with a rich support for QoS.

[13, 18] used DDS in defense system to improve interoperability, high combat survivability & maintainability, and the high performance distributed communication, tactical information management [8].

## 4. Hardware configuration and used testbed

We use in our Labs a real existing Simulation Platform called PLATSIM (see figure 5). Basically, PLATSIM is a distributed interactive simulation platform where users interact with each other over Publish/subscribe middleware. Both DDS and HLA middleware were configured separately to provide human-in-the-loop simulation. The considered application, in the civil domain, is used for remote education in driving schools.

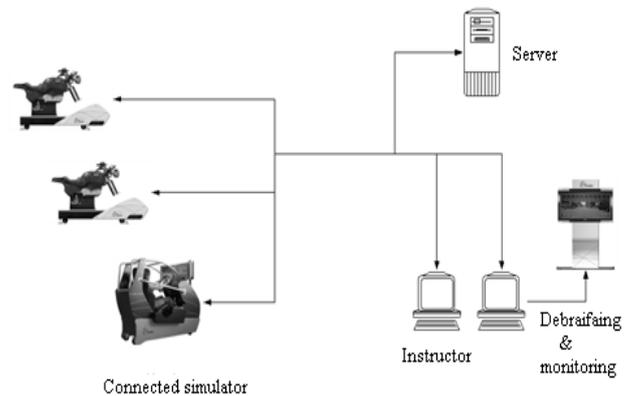

Figure 5: Platsim Hardware testbed

A simulator allows visual modeling of vehicle driven, evaluation of individual pilot's actions, speech synthesis, speech recognition, and recognition of gestures.

The instructor includes the preparation of scenarios, interactivity suitable for evaluation and action on the collective and/or individuals and debriefing (replay) scenarios.

The server supports the implementation of scenarios depending upon instructor or a current event, the calculation of the surrounding traffic , the assessment of collective action, the analysis of symbolic information (voice and gestures) and the calculation of impacts (traffic environment).

In order to measure the latency, a reliable reference time standard was needed. The testbed used in the simulation is synchronized using the Network Time Protocol (NTP) [10]. An NTP server (see figure 6) was used to synchronize all federates with the same reference clock.

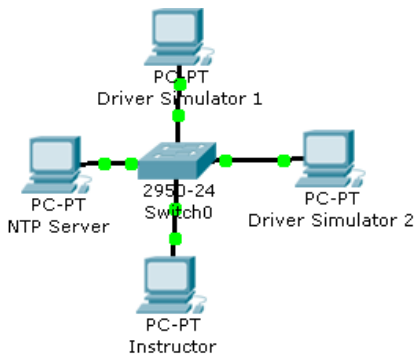

Figure 6: testbed used for the Benchmark

To measure the one way delay in HLA, a simple Federation Object Model (FOM) (see figure 1) and two federates were developed. The FOM consists on several data attributes used in real human-in-the loop simulator. The sender federate publishes its data using multicast transport service. Measurements were stored in trace files and then analyzed separately. The receiver federate subscribes to the object classes and interaction classes. Also, traces files containing time and data information reference of both the publisher and the subscriber were generated.

The RTI under tests was MAK Real Time RTI [11]. It is currently available free of charge, but it can run only between two federates.

In the DDS based simulation, the network latency has been measured using two participant processes. The subscription process consists on an operation that associates a subscriber to its matching publisher, as shown in figure 7.

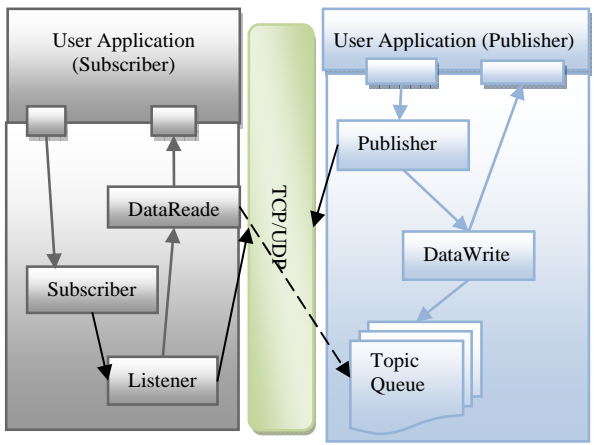

Figure 7 : Model of the DDS simulation

In fact, DDS uses a Real-Time Publish/Subscribe wire protocol (RTPS) to provide a high data rate communication. The RTPS protocol targeted the industrial automation community and then was developed to support the requirements of data distribution systems. It is designed to be able to run over multicast and connectionless best-effort transport protocols like UDP/IP. The RTPS protocol is build on top of UDP (RTP like protocol).

In addition, the subscription process was chosen a topic based subscription. Thus, each data type used by DDS is defined using IDL. The IDL file (see figure 8) is used to identify the data types that DDS processes. These data types are processed by RTIIDLGEN compiler to generate code necessary for transmitting these types with DDS.

```
struct Climat {

 unsigned long    key;
 float climatDistVisi;
 float climatHeure;
 long  climatSport;
 long  climatHorizon;
 float rainDensity;
 float rainSize;
 float wiperAngle;
};
```

Figure 8: IDL structure for the benchmark

Since DDS allows the use of different QoS levels, we need to define how these QoS levels can be guaranteed. In fact, the matching process for QoS guarantee uses a requested/offered (RxO) model. The requested QoS by the subscriber DataReader is less than the offered QoS provided by the publisher DataWriter. The Topic was adjusted to use the same QoS as the DataReader and the DataWriter (see figure 7). The default QoS setting was applied to both the publisher and subscriber: the reliability QoS default settings are best-effort: DDS will send data samples only once to DataReaders. No effort or resources are spent to track whether or not sent sample are received. Data samples may be lost.

## 5. Simulation & discussions

This section analyzes the results of our benchmark conducted using a simulation platform. A set of tests with various configurations has been designed to measure the effects of network latency and jitter and establish performance comparison between HLA-RTI middleware and DDS infrastructure and compare how well HLA and DDS satisfy requirements with respect of the data payload. Figure 9 compare the latency budget results for single node running on HLA and DDS based simulation. The latency budget specifies the maximum acceptable delay from the time the data is written until the data is inserted in the receiver's application cache and the receiving application is notified.

For the inspection of figure 9, we observe that both DDS and HLA are well suited for real-time distributed application. These applications require efficient data collection and delivery. Only minimal delays should be introduced.
The Publish/Subscribe middleware presented here greatly reduces the overhead required to send data over the network compared to client-server architecture. DDA and HLA often care about the determinism of delivering periodic data as well as latency of delivering data.

Occasional subscription requests at low bandwidth replace high bandwidth client requests. In archetypal distributed application, the bandwidth required for distributed nodes even for the same data are quite different.

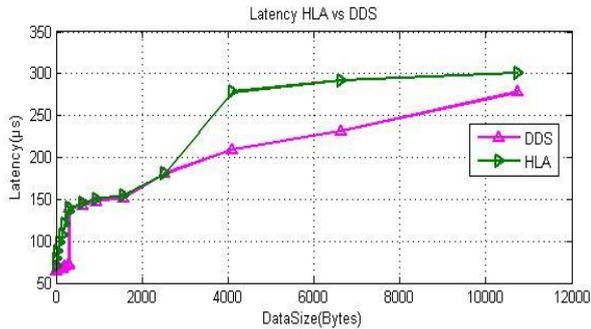

Figure 9: Point-to-point node Latency

Figure 10 compares the jitter results for the same experiments. DDS is match up to provide somewhat better performance than HLA.

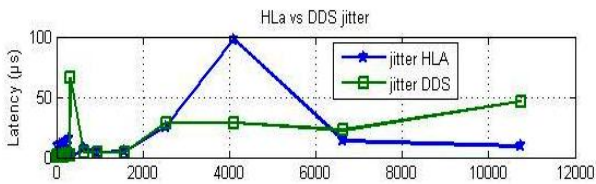

Figure 10: Point-to-point node jitter

Indeed, DDS has good overall performance expected by the most DIS applications. Table 1 strengthens this finding: sample mean and sample median are used to measure the location and the dispersion of the network latency budget and the jitter. Although these results the HLA and DDS latency characteristics are very close. Thus, the close difference between the HLA and DDS middleware may appears when evaluating their performance. But this does not prevent getting very good performance in several distributed applications.

Table 1: Statistic elements for HLA and DDS

|  | HLA | | DDS | |
|---|---|---|---|---|
|  | Latency (µs) | Jitter (µs) | Latency (µs) | Jitter (µs) |
| Mean | 154,87 | 14,13 | 126,60 | 13,36 |
| Median | 138,93 | 9,07 | 106,00 | 3,49 |

Table 2 presents the performance of data transmission vs. the throughput. In such cases, throughput has increased several folds, approaching much more closely the physical limitations of the underlying network transport.

Table 2: Throughput (Mb/s) vs. Packet size (Byte) For HLA and DDS

| Packet size | 10 | 100 | 1000 | 5000 |
|---|---|---|---|---|
| HLA1516 | 2 | 30 | 128 | 350 |
| DDS | 6 | 40 | 112 | 800 |

In addition, both HLA and DDS use a dynamic adjustment to maximize the throughput, and perform the reliability in response to the current network conditions.

An important advantage of HLA and DDS is that they can offer reliability on top of wide variety of transports, including reliable protocols (TCP), unreliable networks (UDP), multicast capable protocol (RAMP, Simple UDP Multicast).

HLA accomplishes this by the capability of implementing RTP/RTCP protocol to ensure more flexible support to exchanged data. In other hand, DDS achieves its performance by employing (optional) a reliable protocol that monitors the liveliness of the link called Real-Time Publish Subscribe (RTPS) protocol.

RTPS is highly configurable with a set of parameters that let the application fine-tune its behavior to trade-off latency, responsiveness, liveliness, throughput, and resources utilization.
From the above tables, it should be noted that DDS improve much more performance than HLA. This is due to specific characteristics of each middleware solution.

Among the HLA services enumerated in Section 2, the Time Management Service is not supported within DDS. It was primarily specified for Parallel and Distributed Architectures. The HLA standardized API specify a save/restore services which ensure the creation of synchronization point between distributed systems to offer more consistence and reliability to applications.

In another hand, the HLA-RTI allows applications to choose the degree to which it participates in time management. The Time Management has to ensure that events are delivered to applications in correct order, but the sequence in which events arrive at the remote application cannot be guaranteed. Events do not arrive in the order of cause and effect relation.

The purpose of both HLA and DDS is to facilitate the efficient use of distributed data in large scale distributed systems; they attempt to unify the common practice of several specific vendor implementations to allow the interoperability and the reusability of existing application. HLA and DDS architectures are common in some regards: using publish/ subscribe paradigm and offering message oriented decentralized communication model. Data dissemination between producer and consumer allows one-to-one, one-to-many, many-to-one and many-to-many communications.

It is significant to note that the next generation of DIS applications requires not only latency management, but also they need advanced end-to-end QoS guarantee on which DDS QoS services can be mapped.

## 6. Conclusion

This paper introduced two middleware architectures based on Publish/Subscribe model and addressing the specific requirements of time-critical, data-critical and large scale distributed interactive systems. HLA is general purpose architecture which aims to interoperate very high number of distributed systems, and DDS is data-centric communication framework with a rich set of QoS Policies, address the challenge of information exchange in high performance communication systems.

DDS service is particularly targeting real-time application, shows its performance when used in another parallel domain which has its specific standards like HLA.

We conducted a benchmark to compare the performance of both DDS and HLA implementation for point-to-point latency budget, jitter and bandwidth utilization in distributed interactive simulation (DIS) application.

Based on our results and experience in distributed interactive simulation and real-time application we learned that DDS holds great promise for DIS applications regarding its high performance compared to HLA.

Future work will look into how to provide QoS guarantee in wide area networks using advanced infrastructure for Next Generation Network architecture that builds, uses and manages end-to-end QoS across different administrative domains and heterogeneous networks.

## 7. Acknowledgement

This research is supported by the French FUI-DGE (Single Inter-Ministerial Fund of the Directorate General for Enterprise) program within the network simulation Platform (PLATSIM).

## Authors Biographies

**AKRAM HAKIRI** is a Ph. D student in the University of Toulouse and researcher in the LAAS-CNRS French research Labs, Toulouse-France. He has his master degree from the University of Paul Sabatier in Toulouse, France and he worked in Wireless Sensor Networks for spatial and Aeronautic systems. He is also an engineer in computer science and automatics from the National Institute of Applied science and Technology (INSAT) in Tunisia.

**PASCAL BERTHOU** is an Associate Professor in computer science in the University of Toulouse and researcher in the LAAS-CNRS French research Labs, Toulouse-France. He worked in network support for the distributed interactive simulation, wireless sensor networks, multi-network communication architecture and multimedia applications over broadband satellite systems.

**THIERRY GAYRAUD is** Full Professor in the University of Science, Toulouse III, France and researcher in the LAAS-CNRS French research Labs, Toulouse-France. His research interests are sensor networks, QoS in satellite communication system and QoS for distributed interactive simulation application.